\documentclass[a4paper]{jpconf}
\usepackage{graphicx}
\usepackage{hyperref}
\hypersetup{pdftex,colorlinks=true,linkcolor=blue,citecolor=blue,menucolor=black,urlcolor=blue,filecolor=blue}
\begin{document}

\title{Holographic black hole engineering at finite baryon chemical potential}

\author{Romulo Rougemont}

\address{\small{\it Instituto de F\'{i}sica, Universidade de S\~{a}o Paulo, Rua do Mat\~{a}o, 1371, Butant\~{a}, CEP 05508-090, S\~{a}o Paulo, SP, Brazil}}

\ead{romulo@if.usp.br}

\begin{abstract}
This is a contribution for the Proceedings of the Conference Hot Quarks 2016, held at South Padre Island, Texas, USA, 12-17 September 2016. I briefly review some thermodynamic and baryon transport results obtained from a bottom-up Einstein-Maxwell-Dilaton holographic model engineered to describe the physics of the quark-gluon plasma at finite temperature and baryon density. The results for the equation of state, baryon susceptibilities, and the curvature of the crossover band are in quantitative agreement with the corresponding lattice QCD results with $2+1$ flavors and physical quark masses. Baryon diffusion is predicted to be suppressed by increasing the baryon chemical potential.
\end{abstract}

\section{Introduction}

\hspace{5mm} Ultrarelativistic heavy ion collisions \cite{expQGP1,expQGP2,expQGP3,expQGP4,expQGP5} constitute the experimental arena where many properties of deconfined QCD matter associated to the formation of a quark-gluon plasma (QGP) \cite{QGP,reviewQGP1,reviewQGP2} are being currently uncovered at the Relativistic Heavy Ion Collider (RHIC) and at the Large Hadron Collider (LHC). One of the most striking features of the QGP produced in such collisions is the strongly coupled nature of the medium close to the hadronization crossover \cite{Aoki:2006we,Borsanyi:2016ksw}. This fact may be attested by the very small value of the ratio between the shear viscosity ($\eta$) and the entropy density ($s$) of the fluid, $\eta/s\approx 0.095$, obtained in hydrodynamic simulations of the spacetime evolution of the QGP simultaneously matching experimental data for different physical observables \cite{Ryu:2015vwa}. Such a small ratio is incompatible with perturbative QCD calculations of $\eta/s$ \cite{Arnold:2000dr,Arnold:2003zc}, but is remarkably close to the value $\eta/s=1/4\pi$ valid for any isotropic and translationally invariant gauge/gravity dual \cite{adscft1,adscft2,adscft3,adscft4} with two derivatives for the metric field in the gravity action \cite{Buchel:2003tz,Kovtun:2004de}. However, this fairly general and robust property of holographic plasmas implies that, in order to attempt to obtain phenomenologically reliable insights for the physics of the QGP, one must take into account the behavior of other physical observables besides $\eta/s$, since many different gauge/gravity duals describing very different strongly correlated quantum fluids share the same value of $\eta/s$.

A largely employed ``toy model'' for the QGP in holographic studies is the $\mathcal{N}=4$ super Yang-Mills (SYM) plasma \cite{solana}, which is one of the simplest and best understood holographic systems at finite temperature. However, a close inspection of its properties reveals that it is a fairly inadequate ``approximation'' of the physics of the real-world QGP. In fact, the SYM plasma is conformal, while the QGP is strongly non-conformal in the crossover region, which is precisely the region where, in principle, holographic plasmas could be useful for real-world phenomenology, since it is in this region where the QGP is strongly coupled (at very large temperatures the QGP becomes weakly coupled and cannot be described by any holographic system defined in the classical gravity limit of the AdS/CFT correspondence, since gauge/gravity duals are known to lack asymptotic freedom, displaying a strongly coupled instead of a trivial ultraviolet fixed point). As a consequence, the equation of state (EoS) of the SYM plasma is completely different from lattice QCD results \cite{Borsanyi:2012cr}, with the most striking difference being imprinted in the trace anomaly of the energy-momentum tensor, which displays a marked peak at the crossover in QCD, while being identically zero for the SYM plasma. Moreover, in what regards transport properties of the QGP, it has been recently pointed out by hydrodynamic simulations \cite{Ryu:2015vwa,Noronha-Hostler:2013gga,Noronha-Hostler:2014dqa,Bernhard:2016tnd} that a nonzero bulk viscosity ($\zeta$) with a peak for the ratio $\zeta/s$ at the crossover seems to be required in order to simultaneously describe different experimental data from heavy ion collisions, which is completely different from the vanishing bulk viscosity of the SYM plasma. Therefore, in order to apply holography to obtain useful results for the real-world QGP, one needs to resort to a different kind of gauge/gravity dual. In this regard, it was originally proposed by Steven Gubser and different collaborators in Refs. \cite{Gubser:2008ny,DeWolfe:2010he} a non-conformal bottom-up Einstein-Maxwell-Dilaton (EMD) model which is able to match the lattice QCD EoS at finite temperature ($T$) and zero baryon chemical potential ($\mu_B$). In Refs. \cite{Finazzo:2014cna,Rougemont:2015wca,Rougemont:2015ona,Finazzo:2015xwa} this construction was further refined and the holographic EoS as well as many transport coefficients were calculated in the $(T,\mu_B)$ plane. In Refs. \cite{Rougemont:2015oea,Finazzo:2016mhm,Critelli:2016cvq} an anisotropic version of the EMD model was proposed to describe the physics of the QGP in the plane of temperature and magnetic field (at zero chemical potential), with the magnetized EoS and many transport properties being computed. These EMD constructions were found to be in quantitative agreement with a large set of observables calculated on the lattice, both at finite $\mu_B$ \cite{Borsanyi:2012cr,Borsanyi:2011sw,Bellwied:2015lba,Bonati:2015bha,Bellwied:2015rza} and finite magnetic field \cite{Bali:2014kia,Bruckmann:2013oba,Endrodi:2015oba,Bazavov:2016uvm}.

\section{Thermodynamics and baryon transport in the $(T,\mu_B)$ plane}

\begin{figure*}
\begin{center}
\begin{tabular}{c}
\includegraphics[width=0.40\textwidth]{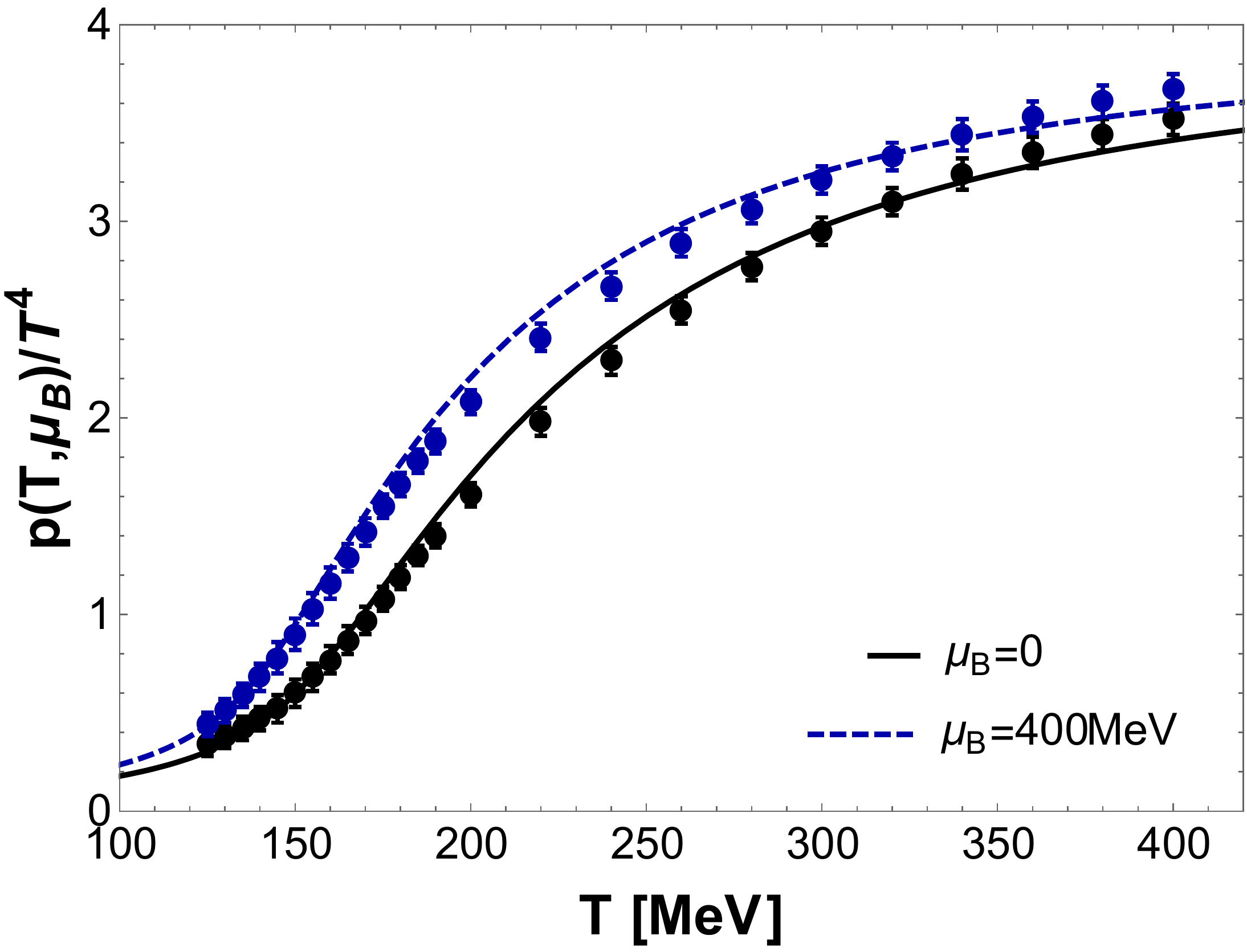}
\end{tabular}
\begin{tabular}{c}
\includegraphics[width=0.40\textwidth]{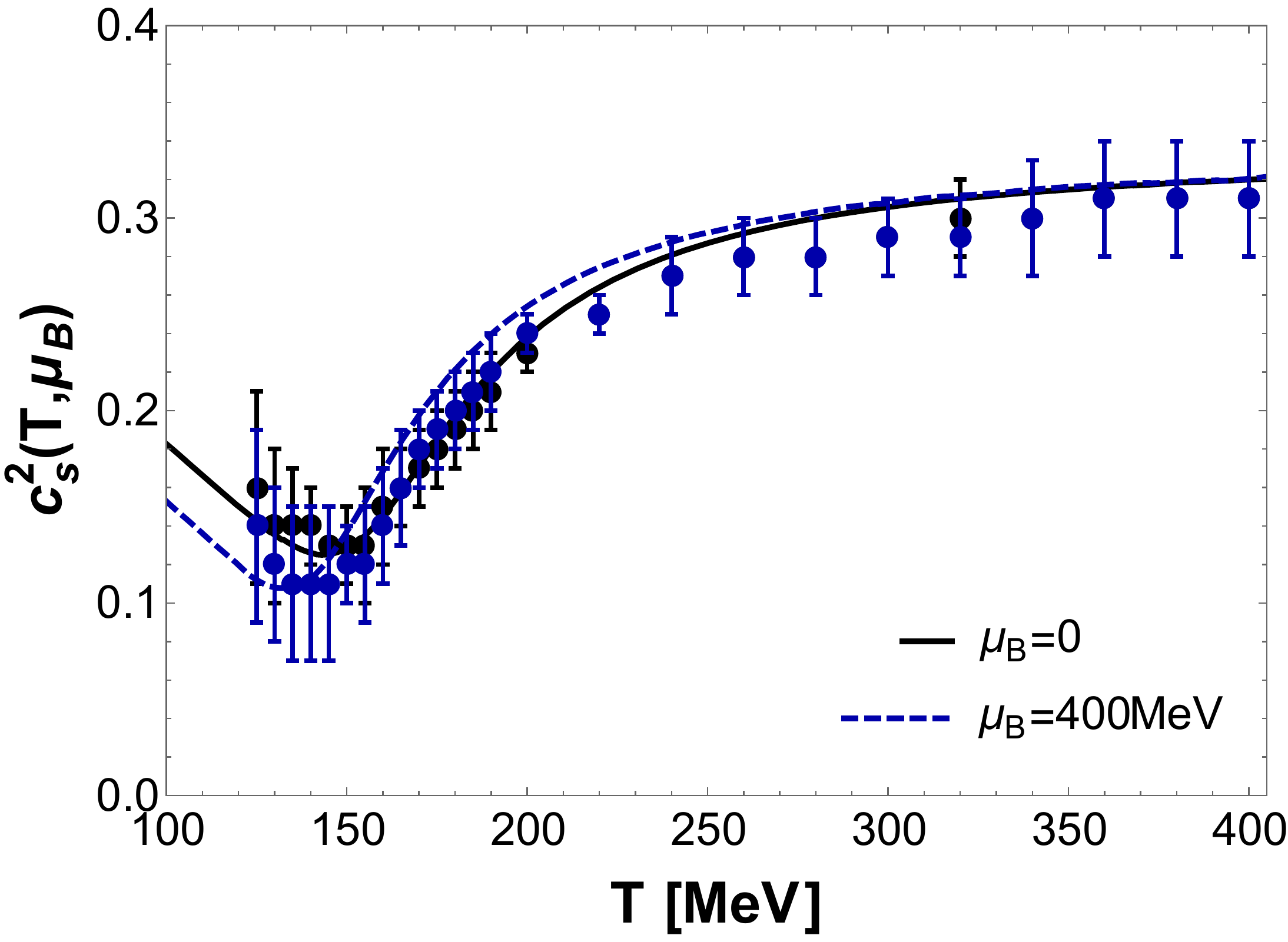}
\end{tabular}
\begin{tabular}{c}
\includegraphics[width=0.40\textwidth]{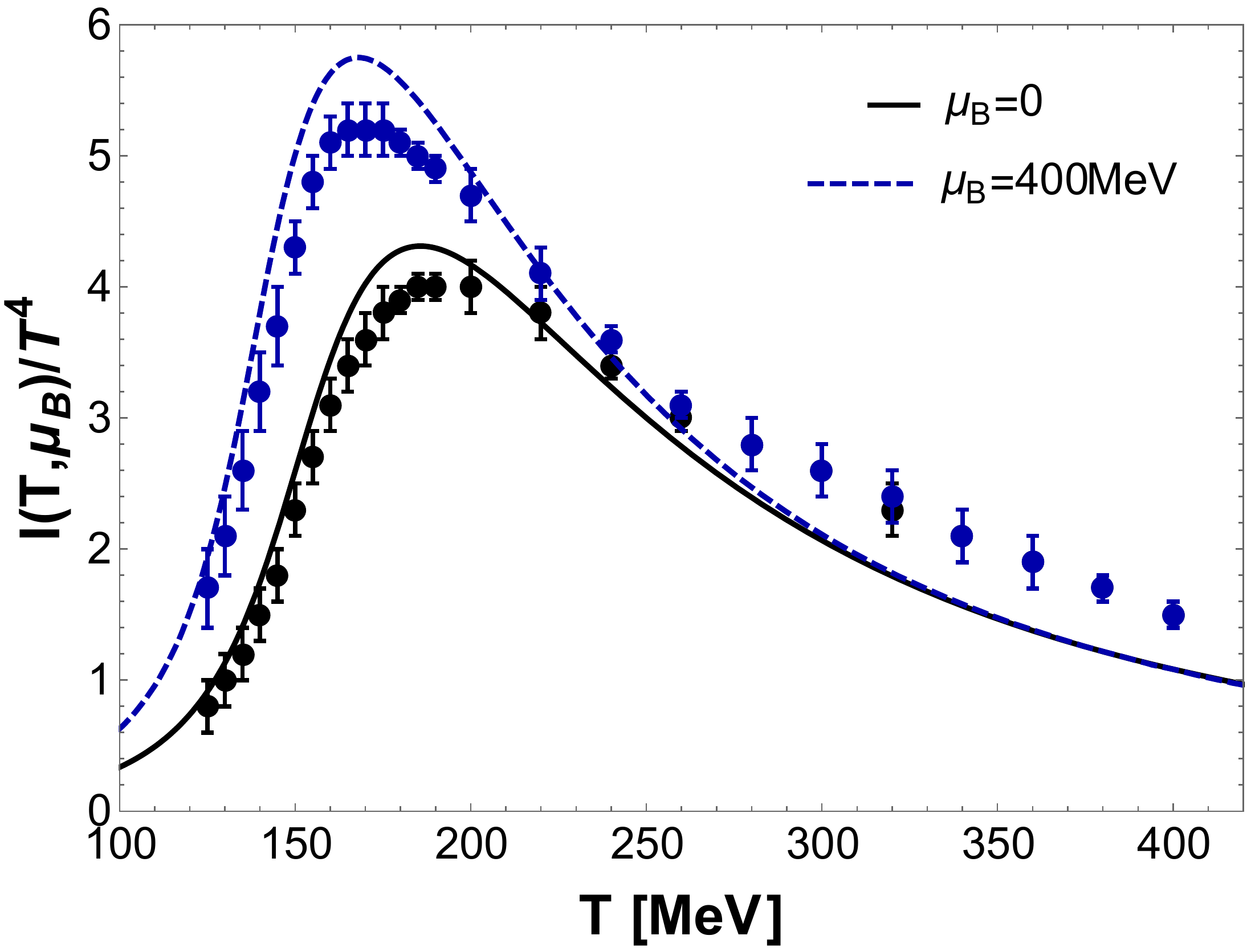}
\end{tabular}
\begin{tabular}{c}
\includegraphics[width=0.42\textwidth]{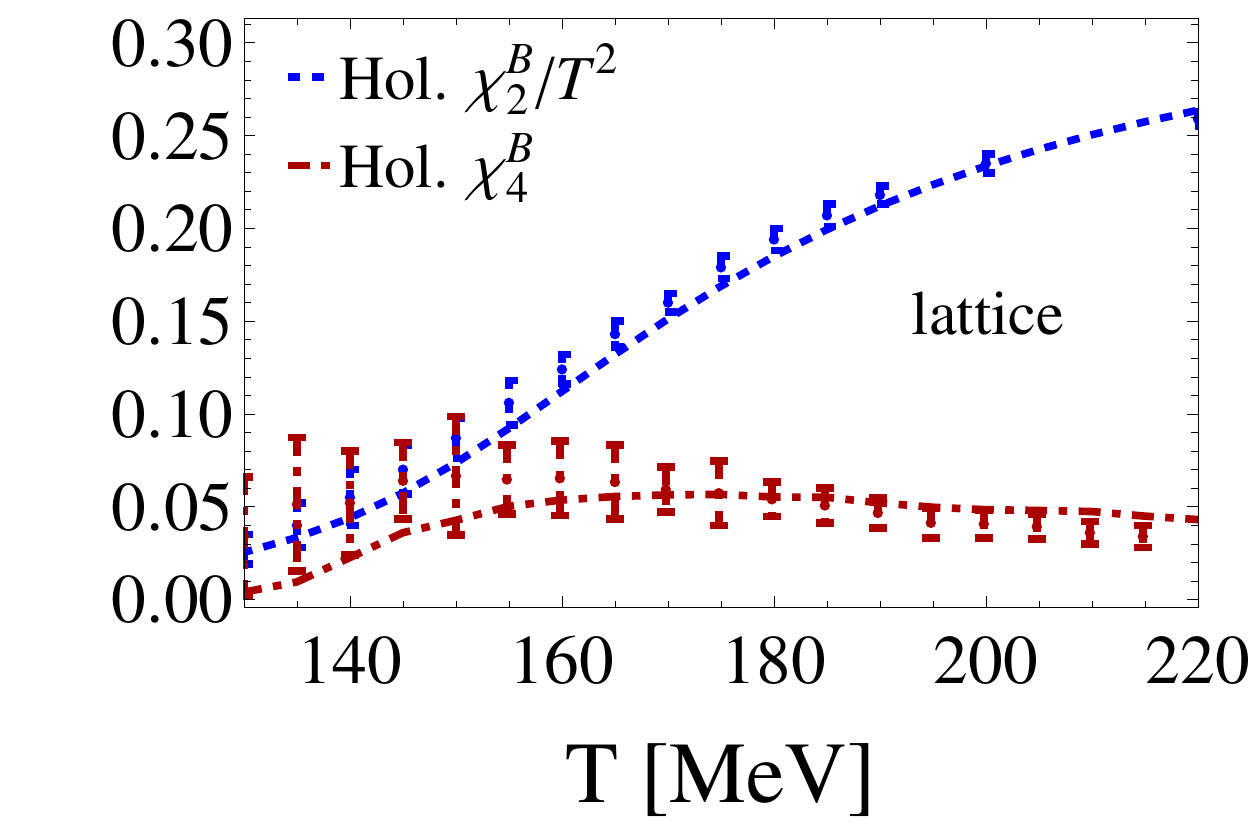}
\end{tabular}
\end{center}
\caption{\label{fig1} {\small \textit{Top left:} pressure. \textit{Top right:} speed of sound squared. \textit{Bottom left:} trace anomaly of the energy-momentum tensor. \textit{Bottom right:} second and fourth order baryon susceptibilities at $\mu_B=0$.}}
\end{figure*}

\hspace{5mm} In Refs. \cite{Rougemont:2015wca,Rougemont:2015ona} the EoS and baryon susceptibilities were calculated in the holographic EMD model and compared to lattice QCD data from Refs. \cite{Borsanyi:2012cr,Borsanyi:2011sw,Bellwied:2015lba}, as shown in Fig. \ref{fig1}. Furthermore, in Ref. \cite{Rougemont:2015wca}, it was also obtained the curvature of the crossover band in the EMD model, $\kappa_{\textrm{EMD}}\approx 0.013$, which is in quantitative agreement with the lattice results $\kappa_{\textrm{latt.}}^{\textrm{(I)}}=0.0135(20)$ from Ref. \cite{Bonati:2015bha} and $\kappa_{\textrm{latt.}}^{\textrm{(II)}}=0.0149(21)$ from Ref. \cite{Bellwied:2015rza}.

\begin{figure*}
\begin{center}
\begin{tabular}{c}
\includegraphics[width=0.35\textwidth]{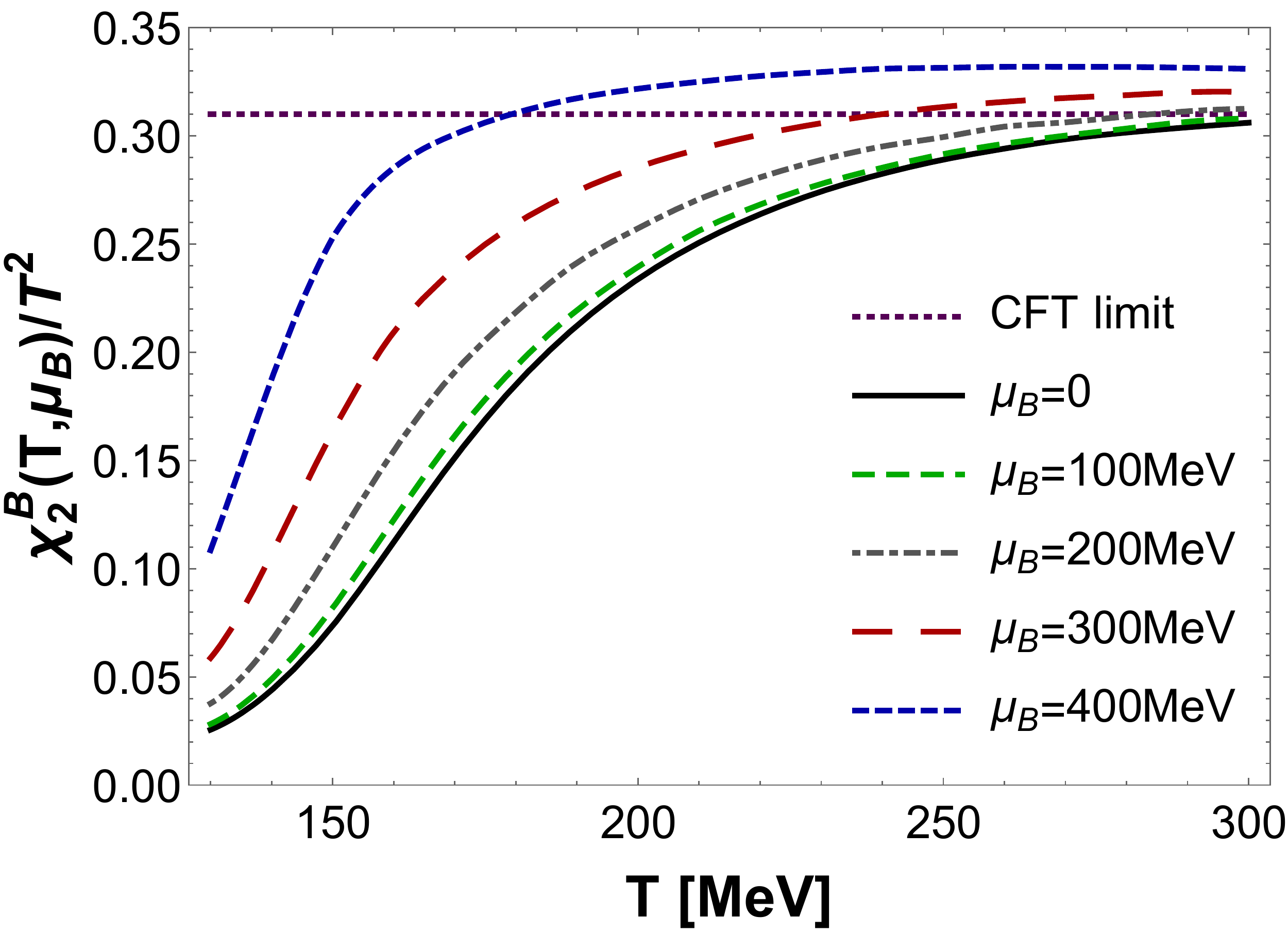} 
\end{tabular}
\begin{tabular}{c}
\includegraphics[width=0.35\textwidth]{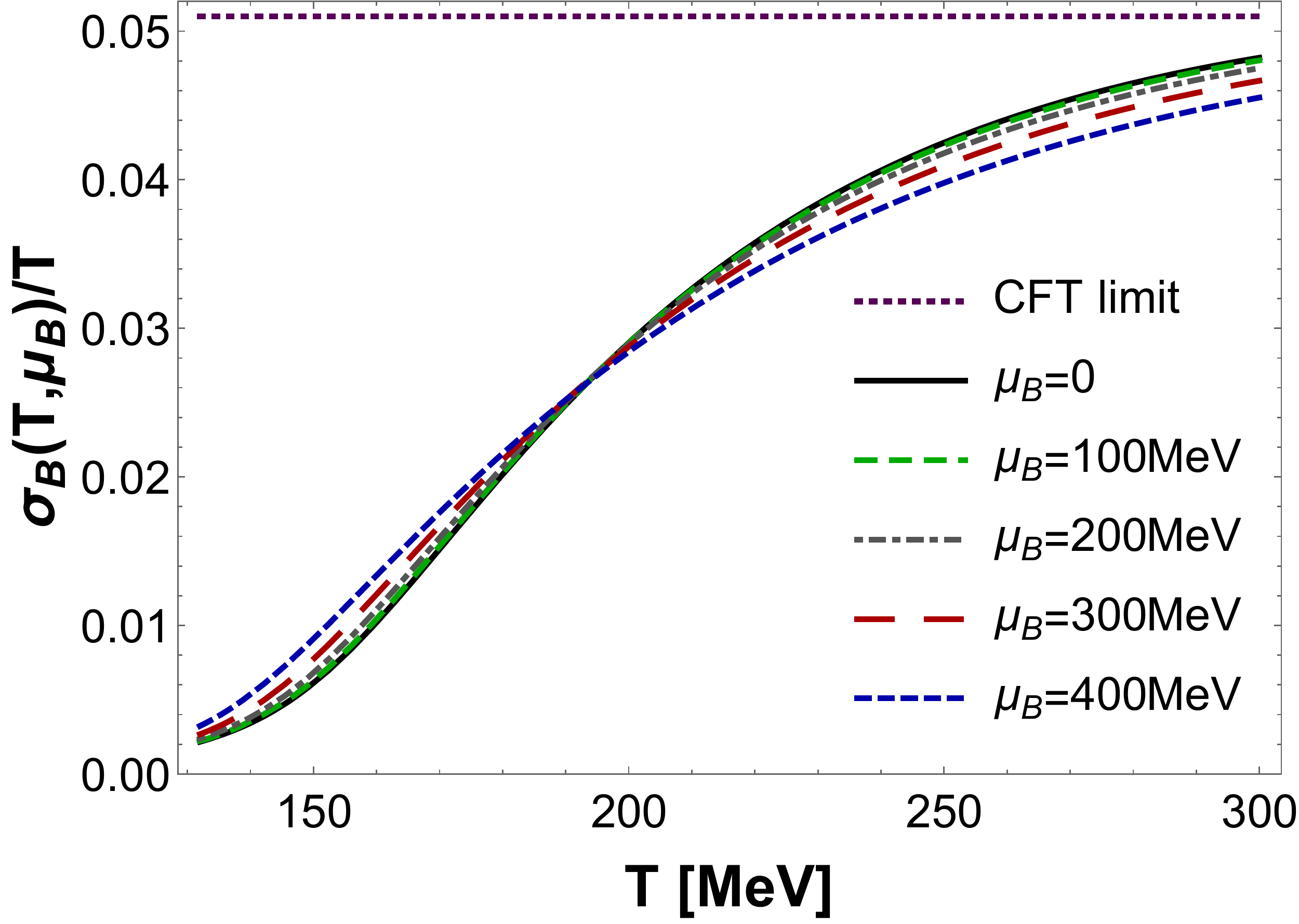} 
\end{tabular}
\begin{tabular}{c}
\includegraphics[width=0.35\textwidth]{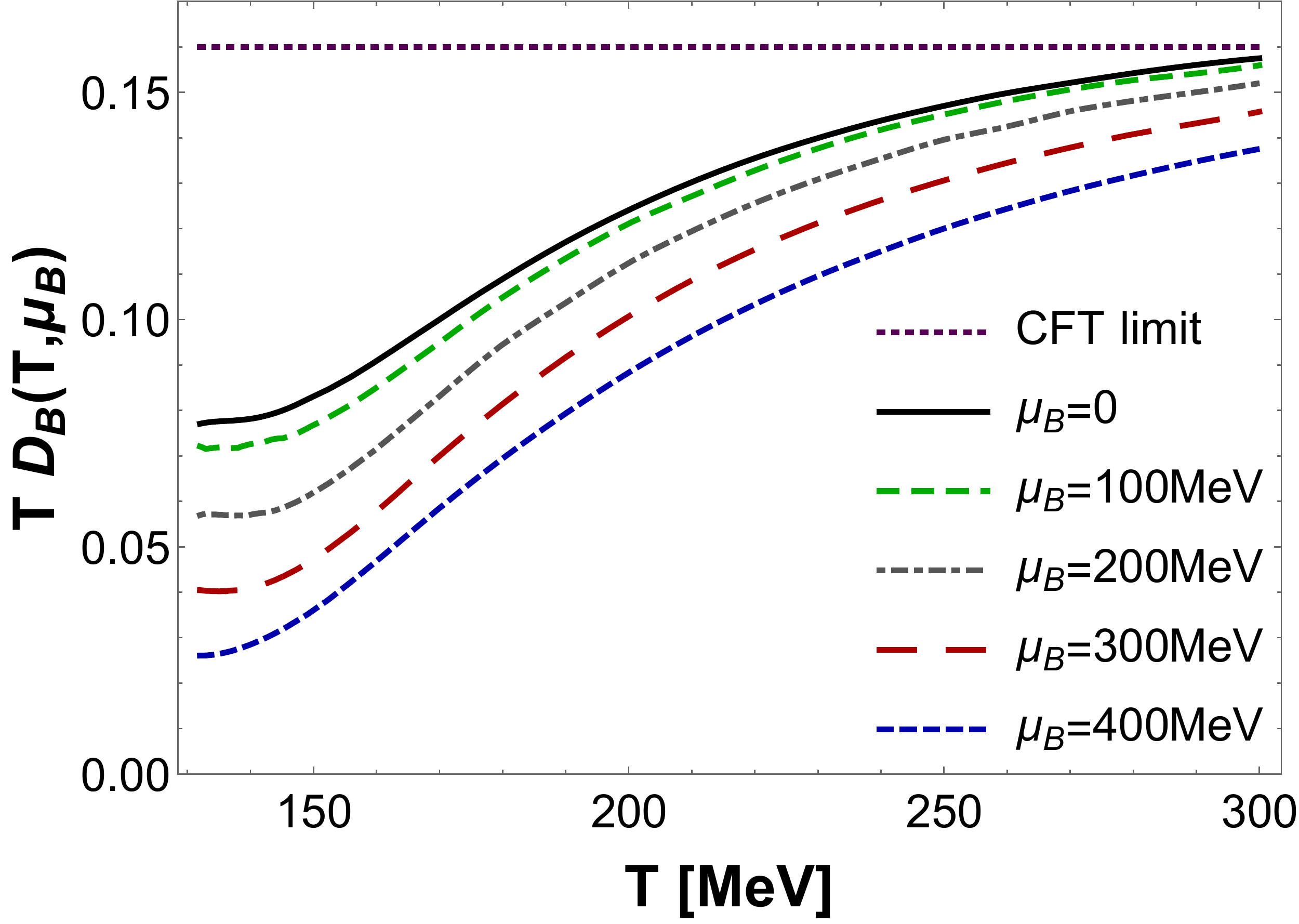} 
\end{tabular}
\end{center}
\caption{\label{fig2} {\small \textit{Top left:} baryon susceptibility. \textit{Top right:} baryon DC conductivity. \textit{Bottom:} baryon diffusion constant. The corresponding conformal limits reached at $T\gg\mu_B$ are also displayed as straight lines in these plots.}}
\end{figure*}

The amount of simultaneous agreement with lattice QCD results for different physical observables, both at zero and nonzero $\mu_B$, together with the \emph{sine qua non} condition for a \emph{bona fide} description of the strongly coupled QGP benchmarked by its nearly perfect fluidity encoded in a small value of $\eta/s$, which is naturally embedded in any gauge/gravity dual, makes the EMD model of Refs. \cite{Rougemont:2015wca,Rougemont:2015ona} a promising tool to make new predictions about the behavior of physical observables which are very difficult to access via first principle lattice QCD calculations, like transport coefficients.

In Ref. \cite{Rougemont:2015ona} the second order baryon susceptibility, the baryon DC conductivity and the baryon diffusion constant, which controls the fluid response to inhomogeneities in the baryon charge density, were calculated as functions of $T$ and $\mu_B$ in the EMD model, as displayed in Fig. \ref{fig2}. The diffusion of baryon charge is predicted to be suppressed as one increases the baryon density, which is compatible with the existence of a critical point in the EMD model for $\mu_B>400$ MeV, as the baryon diffusion is expected to vanish at the critical point \cite{Son:2004iv}. The precise location of the critical point of this holographic EMD model will be determined elsewhere.

\section*{Acknowledgements}

I thank J. Noronha for useful comments on this manuscript and also acknowledge financial support by the S\~{a}o Paulo Research Foundation (FAPESP) under the grant no. 2013/04036-0.

\end{document}